\documentclass[twoside,12pt,fleqn]{article}
\usepackage{amsmath}%\usepackage{astron}
\usepackage{epsfig}\usepackage{psfrag}\usepackage{subfigure}
\usepackage{tabularx}\usepackage{rotating}
\numberwithin{equation}{section}

\makeatletter
\renewcommand{\@thesubfigure}{(\alph{subfigure})}
\renewcommand{\p@subfigure}{}
\renewcommand{\@makecaption}[2]{%  
\vskip 10\p@   \setbox\@tempboxa\hbox{#1.\space#2}%  
\ifdim \wd\@tempboxa >\hsize       #1.\space#2\par     \else       \hbox to\hsize{\hfil\box\@tempboxa\hfil}%  
\fi}
\def\citefull{\def\astroncite##1##2{##1 
##2}\@internalcite}\def\@cite#1#2{#1\if@tempswa  #2\fi}

\def\@fnsymbol#1{\ifcase#1\or \mbox{${^{\star}}$}\or
   \dagger\or \ddagger\or
   \S \or \P \or \|\or \mbox{$^{\star\star}$}\or \dagger\dagger
   \or \ddagger\ddagger\or \S\S\or \P\P\or \|\|\else ***
   \fi\relax}
\newcommand{\lstar}{{\mbox{{\large $\star$}}}} 

\renewenvironment{equation}%
    {\@beginparpenalty\predisplaypenalty
     \@endparpenalty\postdisplaypenalty
     \refstepcounter{equation}%
     \trivlist \item[]\leavevmode
       \hb@xt@\linewidth\bgroup $\m@th% $
         \displaystyle }  %\hskip\mathindent}%
        {$\hfil % $
         \displaywidth\linewidth\hbox{\@eqnnum}%
       \egroup
     \endtrivlist}

\renewenvironment{thebibliography}[1]
     {\section*{\refname}%
      \@mkboth{\MakeUppercase\refname}{\MakeUppercase\refname}%
      \list{\@biblabel{\@arabic\c@enumiv}}%
           {\settowidth\labelwidth{\@biblabel{#1}}%
            \setlength{\labelsep}{1pt}
	\setlength{\itemsep}{0pt}
	\setlength{\parsep}{0pt}
            \leftmargin\labelwidth
            \advance\leftmargin\labelsep
            \@openbib@code
            \usecounter{enumiv}%
            \let\p@enumiv\@empty
            \renewcommand\theenumiv{\@arabic\c@enumiv}}%
      \sloppy
      \clubpenalty4000
      \@clubpenalty \clubpenalty
      \widowpenalty4000%
      \sfcode`\.\@m}
     {\def\@noitemerr
       {\@latex@warning{Empty `thebibliography' environment}}%
      \endlist}

\newcommand{\ct}[1]{\mbox{\tiny \cite{#1}}} 

\makeatother
\addtocounter{footnote}{0}
\begin{document}
\noindent {\Large \bf Axisymmetric models for galaxies by equipotential 
and equidensity methods}
\thispagestyle{empty}
\vskip0.5cm
\noindent {\large Zhenglu Jiang$^{1\lstar}$\footnotetext{\footnotesize\noindent $^{\lstar}${This paper was published in Proceedings of the Sixth Conference of China Society, 
Industry and Applied Mathematics, Li D., Zhang X., Yuan Y., eds., Research Information Ltd., Hertfordshire, p. 79-83, 2002}.
E-mail: 
mcsjzl@zsu.edu.cn (ZJ);
dyfang@mail.hz.zj.cn (DF); moss@ma.man.ac.uk (DM).}, 
Daoyuan Fang$^{2\lstar}$ and David Moss$^{3\lstar}$}
\newline{\footnotesize $^1$Department of Mathematics, Zhongshan 
University,
Guangzhou 510275, P.~R.~China}
\newline{\footnotesize $^2$Department of Mathematics,
Zhejiang University, Hangzhou 310027, P.~R.~China}
\newline{\footnotesize $^3$Department of Mathematics,
University of Manchester, Manchester M13 9PL, UK}
\vskip0.5cm

%\vskip0.6cm

\begin{abstract}
In this paper we outline equipotential and equidensity methods of constructing axisymmetric models for galaxies. 
The former method defines equipotentials, from which the corresponding densities of the galaxy models 
can be obtained using Poisson's equation; the latter defines the equidensity surfaces of the galaxy models directly.
\end{abstract}
\vskip0.5cm

\section{Introduction}
\label{intro}
Most of the earlier models for galaxies are spherical and purely 
empirical,
 in order to fit the surface brightness of galaxies observed. 
Although spherical models can be used to simulate the
surface brightness of some observed galaxies, galaxies
are mostly not spherical. Thus we construct axisymmetric models
 in order to advance our overall understanding of galaxy formation
and evolution. There are several different ways to construct 
axisymmetric models for galaxies.
One of the basic ideas is to extend existing spherical models to
more general axisymmetric forms. 
The methods of this extension
can usually be classified into two groups.
One is the equipotential method$^{[\ct{ko}]},$ 
which was first introduced by
Kuzmin$^{[\ct{k53}, \ct{k}]}$ and developed by Toomre$^{[\ct{t62}]},$
Miyamoto \& Nagai$^{[\ct{mn}]},$ Satoh$^{[\ct{s}]},$ 
Kutuzov \& Ossipkov$^{[\ct{ko81}, \ct{ko86}]},$ 
Evans$^{[\ct{e},\ct{edl}]},$ Jiang$^{[\ct{jiang}]},$  
Ossipkov \& Binney$^{[\ct{ob}]},$
Jiang \& Moss$^{[\ct{jm}]}$ and Jiang et~al.$^{[\ct{jflm}]}.$ This 
device is 
to define equipotentials from which the corresponding
densities of galaxy models can be derived using Poisson's equation. The
significance of this device is that it allows flattening to be achieved 
while
maintaining a simple form of the potential. In particular,
comparing the circular velocity profile
generated from the potential of the model 
with the rotation curve of any very flattened galaxy, we can easily
determine whether the model may be applied to describe the galaxy.
An alternative is the equidensity method, which 
defines the equidensity surfaces of the galaxy models directly; this group includes the
flattened $\gamma$-models$^{[\ct{dg}]}.$  The advantage of this method is 
that
the projected surface densities of the models can be easily derived from 
their
equidensity surfaces, allowing determination of whether these surface densities 
can be used to fit the surface brightness of observed galaxies.

\section{The Equipotential Method}
\label{epm}
Although the construction of galaxy models by the equipotential method has been outlined
by Kutuzov \& Ossipkov$^{[\ct{ko}]},$  we introduce below
our rather different ideas about the method. Most of these ideas have been given by
Jiang$^{[\ct{thesis}]}.$  The equipotential method of constructing
axisymmetric models for galaxies is currently being developed further.
Up to now, the method can be roughly divided into three groups.
The first is to use the axisymmetric radius of $\sqrt{R^2+(a+|z|)^2}$ in 
place of the spherical radius $r=\sqrt{R^2+z^2}$ in the potentials of
the spherical models, where $a>0.$
This was originally introduced by Kuzmin$^{[\ct{k53}, \ct{k}]}.$ 
It is not directly relevant to elliptical galaxies,
and it is used to model discs.
The physical significance of this device is that at the point $(R,-|z|)$ 
below Kuzmin's disc, the potential of Kuzmin's models$^{[\ct{k}]}$  
\begin{equation}
\Phi(R,z)=-\frac{GM}{\sqrt{R^2+(a+|z|)^2}}
\label{kup}
\end{equation}
 is identical with that of a point mass located at distance
$a$ above the centre of the disc.
This implies that $\nabla^2\Phi(R,z)=0$ for $z\not=0.$
As $a$ tends to zero, Kuzmin's models becomes
point mass models.
The second is to use the axisymmetric radius $\sqrt{R^2+(z/q)^2}$ in
place of the spherical radius $r$ of the potentials of
spherical models,
where $q$ is the axial ratio,
 as in Binney's logarithmic models$^{[\ct{bt}]}.$ 
Evans's power-law models$^{[\ct{e}]}$ are given by a simple
power-law potential
\begin{equation}
\Phi(m)=-\frac{\psi_c R_c^p}{(R_c^2+m^2)^{p/2}}, \hbox{   }
 p>0. \label{evans}
\end{equation}
where $m$ is the axisymmetric radius given by $m^2=R^2+(z/q)^2,$
$\psi_c$ is
the potential at the origin and $R_c$ is the scale radius. 
In fact, Evans's power-law models can be regarded as
extensions of Plummer's spherical models$^{[\ct{p11}]}$,
 since (\ref{evans}) degenerates to
the spherical Plummer models when $q=p=1.$
The third is to use
the axisymmetric radius $\sqrt{R^2+(\sqrt{z^2+c^2}+d)^2}$  in
place of the spherical radius $r$ of the potentials of
the spherical models.
 This device was first introduced by Miyamoto and Nagai$^{[\ct{mn}]}$ 
for the Plummer potential, when they constructed
the following potential-density pair
\begin{equation}
 \Phi(R,z)=-\frac{GM}{\sqrt{R^2+(\sqrt{z^2+c^2}+d)^2}},
\label{plukup}
\end{equation}
and
\begin{equation}
\rho(R,z)=\left(\frac{c^2M}{4\pi}\right)\frac{dR^2+(d+3\sqrt{z^2+c^2})
(d+\sqrt{z^2+c^2})^2}{[R^2+(d+\sqrt{z^2+c^2})^2]^{5/2}(z^2+c^2)^{3/2}}.
\label{plukud}
\end{equation}
When $c=0,$ (\ref{plukup}) reduces to Kuzimin's potential
mentioned above.
When $d=0,$ (\ref{plukup}) reduces to the spherical potential of 
Plummer.

The first and third devices described above have attracted the attention of many 
outstanding 
investigators. Thus several ideas of constructing axisymmetric
potentials are given and they are the development of these devices.
A new potential-density pair can be obtained
if (\ref{kup}) is differentiated with respect to $a^2.$
Toomre's model $n$ was derived by Toomre$^{[\ct{t62}]}$ by
taking the $(n-1)$st derivative of $(\ref{kup})/a$ with respect to 
$a^2.$
 Similarly, if (\ref{plukup}) is differentiated $n$ times
with respect to $c^2$, Satoh's models$^{[\ct{s}]}$  
can be obtained.

Following Kuzmin's models mentioned above, Kuzmin \& Kutuzov$^{[\ct{kk}]}$ 
constructed a more general axisymmetric potential as follows
\begin{equation}
\Phi(R,z)=-\frac{GM}{(R^2+z^2+a^2+c^2+2\sqrt{a^2c^2+c^2R^2+a^2z^2})^{1/
2}}.
\label{kuku}
\end{equation}
This is a St\"ackel model. Similarly, Evans et~al.$^{[\ct{edl}]}$ 
obtained flattened isochrone models with the potential
\begin{equation}
\Phi(R,z)=-\frac{X+aY+c^2}{Y(X+aY+a^2)},
\label{fim}
\end{equation}
where $X=\sqrt{a^2c^2+c^2R^2+a^2z^2}$ and 
$Y=\sqrt{a^2+c^2+R^2+z^2+2X}.$
In (\ref{kuku}) and (\ref{fim})
 $a$ and $c$ are non-negative length scales.

Recently, oblate Jaffe models for galaxies were given by Jiang$^{[\ct{jiang}]},$ 
using Miyamoto \& Nagai's device. Then
Jiang \& Moss$^{[\ct{jm}]}$ 
constructed a class of prolate Jaffe models, obtained by replacing
the spherical radius $r$ of the potentials of Jaffe's spherical models
with a different axisymmetric radius of 
$\sqrt{(\sqrt{R^2+a^2}+b)^2+z^2},$
where $a$ and $b$ are two positive constants. This
method is similar to that of  Miyamoto \& Nagai.  Furthermore, Jiang 
et~al.$^{[\ct{jflm}]}$ 
gave  more general
flattened Jaffe models with potentials
\begin{equation}
\Phi(R^2,z)=\frac{GM}{r_J}
\ln\left(\frac{\sqrt{(\sqrt{R^2+a^2}+b)^2+(\sqrt{z^2+c^2}+d)^2}}
{\sqrt{(\sqrt{R^2+a^2}+b)^2+(\sqrt{z^2+c^2}+d)^2}+r_J}\right),
\label{(1.3)}
\end{equation}
using a device similar to that of Miyamoto \& Nagai, that is replacing the
spherical radius $r$ of the potentials of Jaffe's models$^{[\ct{j}]}$
with
a more general axisymmetric radius $m$,
where $m=\sqrt{(\sqrt{R^2+a^2}+b)^2+(\sqrt{z^2+c^2}+d)^2},$ 
and $a,b,c$ and $d$ are positive constants.

Since the spherical $\gamma$-model$^{[\ct{d93}, \ct{saha},  \ct{t94}, \ct{vuik}]}$ 
include Jaffe's spherical model, 
we can also define a class of  axisymmetric $\gamma$-models,
replacing the spherical radius $r$ of the potential of 
the spherical $\gamma$-model   
with
the more general axisymmetric radius 
$m$ given above. 

\section{The Equidensity Method}
\label{edm}
The equidensity method directly defines the ellipsoidal equidensity surfaces of the galaxy 
models. 
For example, the axisymmetric radius $\sqrt{R^2+(z/q)^2}$ can be used in
place of the spherical radius $r$ in the densities of the spherical 
models,
where $q$ is the axial ratio, as in Binney's logarithmic models$^{[\ct{bt}]}.$ 
The family of densities of the flattened $\gamma$-model$^{[\ct{dg}]}$ is 

\begin{equation}
\rho(m)=\frac{(3-\gamma)M}{4\pi q}
\frac{am^{-\gamma}}{(m+a)^{4-\gamma}},
\label{dehnenad}
\end{equation}
from the density of the spherical $\gamma$-model 
mentioned in Section \ref{epm},
where $M$ and $a$ are total mass and scale radius,
$m$ is the same as in (\ref{evans}), $q$ is the axial ratio and 
$0\leq\gamma<3.$

The advantage of this method is that
 the projected surface densities of the models are easily
derived.  Thus we can easily determine
whether the surface densities can be used to
fit the surface brightness of the galaxies observed. However,
this method can lead to mathematical difficulties in finding
analytical expressions for the potentials.
%\vskip0.8cm
%\noindent{\bf Acknowledgement}~
{\small
\section*{Acknowledgement}
This work was supported in part by NSFC number 19971077.

}

\end{document}